\documentclass[conference]{IEEEtran}
\IEEEoverridecommandlockouts
\usepackage{cite}
\usepackage{amsmath,amssymb,amsfonts}
\usepackage{algorithmic}
\usepackage{graphicx}
\usepackage{textcomp}
\usepackage{xcolor}
\usepackage{subcaption}
\usepackage{balance}
\usepackage{hyperref} 

\def\BibTeX{{\rm B\kern-.05em{\sc i\kern-.025em b}\kern-.08em
    T\kern-.1667em\lower.7ex\hbox{E}\kern-.125emX}}

\newcommand{\RCAETX}{\textsf{RCA-ETX}}

\newif\ifshowcomments
\showcommentstrue

\ifshowcomments
\newcommand{\mynote}[2]{\textcolor{blue}{\fbox{\bfseries\sffamily\scriptsize#1}}
  \textcolor{blue}{{$/*$\textsf{\emph{#2}}$*/$}}}
\else
\newcommand{\mynote}[2]{}
\fi

\begin{document}

\title{Contact-Aware Opportunistic Data Forwarding in Disconnected LoRaWAN Mobile Networks\\
}

\author{Po-Yu Chen\textsuperscript{\dag},
Laksh Bhatia\textsuperscript{\dag},
Roman Kolcun\textsuperscript{\S},
David Boyle\textsuperscript{\S},
Julie A. McCann\textsuperscript{\dag}\\
\textsuperscript{\dag}\textit{Department of Computing, Imperial College London}\\
\textsuperscript{\S}\textit{Dyson School of Design Engineering, Imperial College London} \\
\textit{
\{po-yu.chen11,laksh.bhatia16,roman.kolcun,david.boyle,j.mccann\}@imperial.ac.uk}
}

\maketitle

\begin{abstract}

LoRaWAN is one of the leading Low Power Wide Area Network (LPWAN) architectures. It was originally designed for systems consisting of static sensor or Internet of Things (IoT) devices and static gateways. It was recently updated to introduce new features such as nano-second timestamps which open up applications to enable LoRaWAN to be adopted for mobile device tracking and localisation. In such mobile scenarios, devices could temporarily lose communication with the gateways because of interference from obstacles or deep fading, causing throughput reduction and delays in data transmission.
To overcome this problem, we propose a new data forwarding scheme. Instead of holding the data until the next contact with gateways, devices can forward their data to nearby devices that have a higher probability of being in contact with gateways.
We propose a new network metric called Real-Time Contact-Aware Expected Transmission Count (RCA-ETX) to model this contact probability in real-time. Without making any assumption on mobility models, this metric exploits data transmission delays to model complex device mobility.
We also extend RCA-ETX with a throughput-optimal stochastic backpressure routing scheme and propose Real-Time Opportunistic Backpressure Collection (ROBC), a protocol to counter the stochastic behaviours resulting from the dynamics associated with mobility. 
To apply our approaches seamlessly to LoRaWAN-enabled devices, we further propose two new LaRaWAN classes, namely Modified Class-C and Queue-based Class-A. Both of them are compatible with LoRaWAN Class-A devices.
Our data-driven experiments, based on the London bus network, show that our approaches can reduce data transmission delays up to $25\%$ and provide a $53\%$ throughput improvement in data transfer performance.
\end{abstract}


\section{Introduction}
\label{sec:intro}

Low Power Wide Area Networks (LPWAN) are a relatively new class of wireless communication systems designed for long-range and low-power performance. There are several LPWAN protocols like SigFox, DASH7, NB-IoT and LoRaWAN\cite{7815384,gao2019towards}. Of these protocols, LoRaWAN is gaining much traction from industrial and academic communities due to its low deployment costs and flexible network-layer protocols. 
LoRaWAN realises relatively long-range wireless communications (e.g. 2~km in urban and 15~km in rural) via sub-GHz frequencies (e.g. 868~MHz in EU and 915~MHz in North America). However, nodes operating in the EU need to adhere to low duty cycle operation ($<1\%$ or $<10\%$)\cite{LoRaWAN} and practical mobile communications scheme must adhere to this.  
 
Much work has been done to improve throughput and reduce delays in LoRaWAN networks generally. Most of these works, however, assume that nodes and gateways are static and do not consider mobility. 
There is a rising class of applications that require low-power mobile solutions to provide new services, e.g. asset tracking.
Semtech, the company that proposed LoRa technology, has also been designing new gateways to support nanosecond precision timestamps for acquiring geolocation information from LoRa-enabled devices for tracking scenarios.
Under mobility, radio channel states vary as locations and environments change over time. Radio channels can be unreliable due to obstacles or deep fading, which can further depend on weather conditions, line-of-sight to base stations being blocked, or the speed of the vehicle hosting the data source. 
The low duty cycling (i.e., 1\% for general data channels) regulated by LoRaWAN specification makes this situation more challenging still. If a device misses its allocated time slot for communication, it needs to wait minutes if not hours before it can upload its data in the next round, thus causing significant delays to data delivery. Consequently, new and/or improved protocols are required to ensure on-demand delivery of information during movement. 

Here, instead of trying to send data via unreliable radio channels to gateways or awaiting for next gateway contact, LoRaWAN devices can send their data to gateways via other nearby devices that have better quality contact with gateways. As the capability of communicating through long-range LoRa is typically more significant than actual device mobility, it is easier to find reliable neighbours for data forwarding than retain unsent data awaiting a good connection with a gateway. By exploiting this observation, we propose two new opportunistic data forwarding approaches which can effectively reduce data delivery delays and improve network throughput.

In the first approach, we introduce a new network metric to drives decision making in terms of the helper devices chosen to forward data through - namely, Real-time Contact Aware Expected Transmission Count (RCA-ETX). 
This new metric can seamlessly illustrate device mobility with packet transmission delay, which is widely accepted in many objective-function-based data forwarding protocols.
Furthermore, we extend RCA-ETX by combining it with a throughput-optimal stochastic backpressure scheme, and propose Real-time Opportunistic Backpressure Collection (ROBC). ROBC can cope better with uncertain links in mobile scenarios, and delivers improved performance compared to RPL with RCA-ETX. Importantly, RCA-ETX and ROBC can operate without prior knowledge of device mobility.
Finally, to ensure our approaches are compatible with other LoRaWAN enabled devices, we propose two new LaRaWAN classes, namely Modified Class-C and Queue-based Class-A, to support device without and with energy constrains, respectively. Both of the proposed classes are compatible with LoRaWAN Class-A, which must be supported by all LoRaWAN enabled devices.

To demonstrate the effectiveness of RCA-ETX and ROBC, we evaluate the performance of RCA-ETX and ROBC with extensive data-driven experiments using the London Bus network \cite{TFLOpen}. Our experimental results show that the solution can effectively reduce delays by up to $25\%$ in both urban and rural areas, while giving a throughput improvement up to $53\%$ with an overhead of up to 2.2 times the number of messages.

The rest of this paper is organised as follows.
Section \ref{sec:relatedwork} discusses related work.
Section \ref{sec:background} introduces the background to this work. Our main contributions, RCA-ETX and ROBC, are explained and proposed in Sections \ref{sec:RCA-ETX} and \ref{sec:ROBC}, respectively. A discussion on practical implementations and performance evaluation is given in Sections \ref{sec:energy_saving} and \ref{sec:evaluation}, and we conclude the paper in Section \ref{sec:conclusion}.

\section{Related Work}
\label{sec:relatedwork}

Opportunistic data forwarding in mobile LoRaWAN networks is a relatively new topic. In this section, we briefly review the recent research on mobile LoRaWAN networks and opportunistic data forwarding in wireless networks.

\subsection{Mobility in LoRaWAN}

Low Power Wide Area Networks (LPWANs) have gained serious traction in the last decade. There are a number of works that compare state-of-art LPWAN protocols (e.g. LoRaWAN, Dash7 and NB-IoT) and evaluate how different stages of the protocol affect performance\cite{7815384,8254509,8407095} and how parameter settings (e.g. spreading factors and listen-before-talk) affect the performance such as delay and data throughput \cite{Liando:2019:KUF:3311822.3293534}. 
In Internet of Mobile Things\cite{8502812}, the authors do a comparative analysis of LoRaWAN, Dash7 and NB-IoT. They discuss mobility considerations and which of the three protocols are suitable for which application areas. Their results show that all of them are useful for certain use-cases and that LoRaWAN is better suited for applications like pallet tracking that require low cost, extended battery life, uplink only communication and long range coverage.

However, mobility also results in new challenges when applied in real-world scenarios. For example, Petjjrvi et al. show that velocity has an observable effect on the packet delivery ratio for speeds higher than 25~km/h due to Doppler effects \cite{doi:10.1177/1550147717699412}, while
Marcelis et al. find the relationship between packet delivery ratio,mobility and distances between devices and gateways \cite{7946866}.
Without reliable links between mobile devices and gateways, opportunistic data forwarding between nearby devices becomes a potential solution.

\subsection{Opportunistic Data Forwarding}

In the areas of Internet of Things (IoT) and Wireless Sensor Networks (WSNs), many recent works have focused on finding fast and reliable ways to forward data given a network where the devices therein are mobile \cite{cai2013routing, condi2014mobile}.
One notable direction focuses on data forwarding in networks consisting of static sensors and mobile sinks \cite{kusy2009predictive,li2011ubiquitous,lee2010whirlpool}. In these networks, resource-constrained sensors are typically disconnected from the Internet and have limited communication capability. These sensors therefore try to maximise the opportunity of forwarding data via nearby mobile devices (e.g. mobile phones) which have a direct connection to the Internet.
Some recent works try to find the better opportunistic routes maximising the probability of successful data forwarding \cite{tahir2017brpl} without making any assumption on mobility. Yang et al. \cite{yang2017practical} proposed a new metric, contact-aware ETX (CA-ETX), to seamlessly merge mobility into classic scheduling protocols (specifically RPL and backpressure routing). However, their work focuses on networks which consist of mobile sinks and static sensors, and it relies on reliable radio channels and high duty-cycle communication between sensors.

Another notable direction focuses on data forwarding in disconnected mobile \textit{ad hoc} networks consisting of static sinks and mobile devices where connections between them are lossy \cite{cai2013routing,pelusi2006oppurtunistic,yuan2015hotspot}.
Some recent studies exploit node contact patterns to reduce delays and improve successful data forwarding\cite{tornell2015dtn}.
These patterns can be generated based on devices' historical motion paths \cite{zhu2013scaling} and social behaviours\cite{zhu2014social,zhou2017predicting}.
\cite{zhou2013exploiting} further takes duty cycles into account by assuming that devices may not be able to communicate with each other at every physical contact when they can communicate. However, these solutions rely heavily on extra information, such as geographical information and social behaviours, which are challenging to obtain and process for resource-constrained LoRa devices.
\section{Background}
\label{sec:background}
In this section, we first provide a formal definition to the Mobile LoRaWAN network with static sinks (MLoRa-SS). We then give a brief description of the LoRaWAN communication specification. Finally, we briefly describe the contact-aware ETX, the foundation of our approach, and the reasons why it cannot be directly applied to MLoRa-SS.

\subsection{System Model of MLoRa-SS}
\label{subsec:MLoRa-SS}

Unlike standard LoRaWAN networks, which consist of only sensor-to-sink links in a star topology, in our Mobile LoRaWAN network with static sinks (MLoRa-SS), we assume devices (i.e., LoRa end-devices therein) may move over time. Take the London bus network as an example: the average bus speeds for different routes varying between 5.4 mph and 23.1 mph. In MLoRa-SS, we aim to provide a solution that allows devices to send their data via other nearby peer devices if their direct connection to their sinks is currently weak or unavailable.

\begin{figure}[!ht]
\begin{center}
\includegraphics*[width=0.95\linewidth]{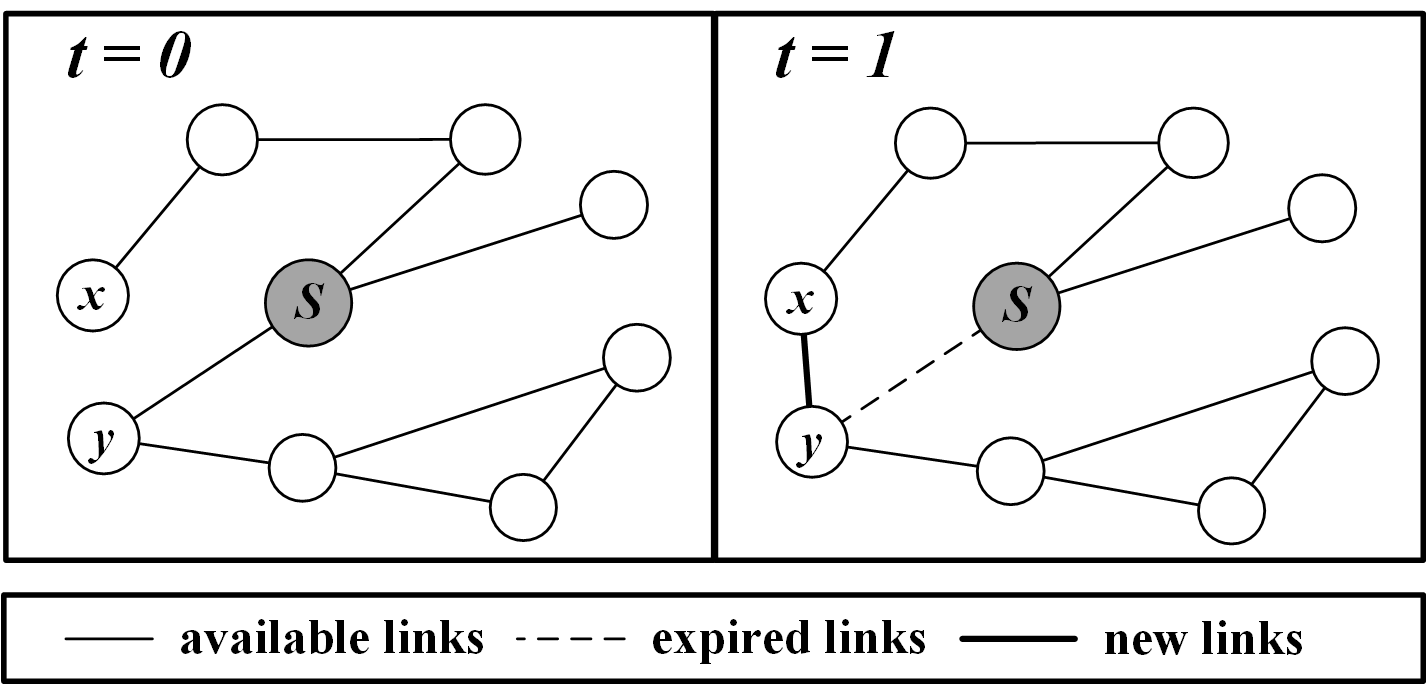}
\caption{Compared to the network at $t=0$, link $(x,y)\in \mathcal{L}$ regains communication (i.e. $c(t)_{x,y} > 0$), whereas device $y$ loses its link with sinks $\mathcal{S}$ (i.e. $c(t)_{x,\mathcal{S}} =0$) at $t=1$.
}
\label{fig:MLoRa-SS_Example}
\vspace{-5mm}
\end {center}
\end{figure}

To better illustrate the dynamics of MLoRa-SS, here we give a formal definition. MLoRa-SS can be modelled as a time-varying weighted graph $G(\mathcal{N},\mathcal{L},\mathbf{C}(t))$. 
In this graph, $\mathcal{N}=\mathcal{D}\cup\mathcal{S}$ denotes all nodes in an MLoRa-SS, where $\mathcal{D}$ denotes the set of all devices (i.e. LoRa end-devices) generating and relaying data packets, and $\mathcal{S}$ is the set of all sinks (i.e. LoRaWAN gateways) collecting data packets from the network. 
$\mathcal{L}=\mathcal{L}_d\cup \mathcal{L}_s$ denotes the set of all possible wireless links.
While $\mathcal{L}_d$ denotes the link between node pairs $\{x,y\} \in \mathcal{D}$, each entry in $\mathcal{L}_s$ represent the virtual links between a node $x\in \mathcal{D}$ and all sinks $s\in \mathcal{S}$.
$\mathbf{C}(t)$ denotes a $|\mathcal{N}|^2$-dimensional matrix representing the channel capacities over all wireless links between nodes at time $t$. 
Each entry $c(t)_{x,y}\in \mathbf{C}(t)$ is dynamic and may change overtime. 
Figure \ref{fig:MLoRa-SS_Example} illustrates an example of $G(\mathcal{N},\mathcal{L},\mathbf{C}(t))$. 

\subsection{LoRaWAN Communication Specification}
\label{subsec:LoRaWAN}

\begin{figure}[!ht]
\begin{center}
\includegraphics*[width=1.0\linewidth]{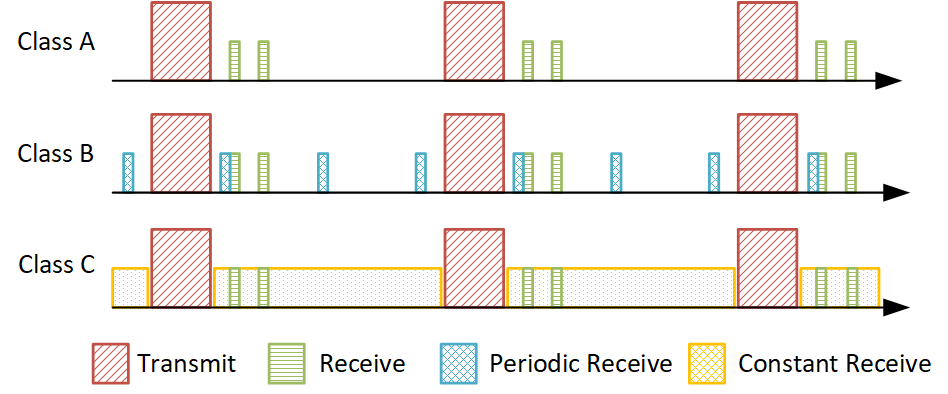}
\caption{LoRaWAN specification defines three devices types. General SRD defines the maximum transmission duty cycle to be $1\%$ (i.e. $l/\Delta t \leq 1\%$).  All devices must implement \textbf{Class A} whereas other two are optional.}
\label{fig:LoRaWAN-Class}
\vspace{-5mm}
\end {center}
\end{figure}

In order to run our solution under MLoRa-SS, we need to ensure our solution is compatible with the LoRaWAN specification. LoRaWAN relies on one-hop broadcasting via LoRa RF technology which is formally regulated. The overall device-to-sink link capacity of each LoRa devices is very low. When a node selects general SRD and SF12/125kHz defined in LoRaWAN specification, the maximum link capacity for such node is only 2.5 bit/s with $<1\%$ duty cycle. Bi-directional communication between a device and a gateway under such constraints are defined in LoRaWAN specification as three device types \cite{LoRaWAN}. All LoRa end-devices must implement Class A, whereas Class B and C are extensions to Class A as illustrated in Figure \ref{fig:LoRaWAN-Class}. 

As can be seen, \textbf{Class A} devices open two receive windows at specific times (i.e. 1s and 2s) after an uplink transmission. A gateway can respond either in the first window or the second window. \textbf{Class B} devices reduce delays in down-link communications extending Class A by adding periodic receive slots. \textbf{Class C} devices extend Class A by keeping the receive windows open unless they are transmitting. It has the lowest delay but is less energy efficient. 

Node to node communication in LoRaWAN can be enabled if sensors overhear packets from other nearby sensors for information exchange and receive window. In this work, LoRaWAN devices use a modified Class-C device. The nodes are always listening to a shared channel to overhear packets from each other and only switch channels to receive an acknowledgement from the gateway. We also explore the idea of Queue-based class-A to reduce the energy used by the nodes. In Queue-based class-A the length of receive slots is decided based on the number of messages in queue.

\subsection{CA-ETX and Challenges when applying in MLoRa-SS}
\label{subsec:challenges}

Contact-Aware ETX (CA-ETX) \cite{yang2017practical} is a network metric that extends Expected Transmission Count (ETX) \cite{de2005high}, which is a widely accepted network metric used by many routing and scheduling protocols in Wireless Sensor Networks (WSN). It takes mobility in WSNs into account when making scheduling and routing decisions and is designed for WSNs with mobile sinks (WSN-MS).

Although CA-ETX provides a novel way to incorporate mobility in the computation of ETX seamlessly, it cannot be directly applied in MLoRa-SS due to the following constraints: 

\begin{enumerate}
    \item \textbf{Opportunistic contact in between sensors.} CA-ETX is designed for WSN-MS, where sinks are mobile while sensors are assumed static. In contrast, in MLoRa-SS, sensors are allowed to move, while sinks are static. Therefore, routing protocols such as RPL or BCP may never converge to a stable routing gradient and consequently cannot provide guarantees such as minimal delay in RPL and maximum throughput in BCP.
    
    \item \textbf{Low maximum duty cycle and limited bandwidth.} As presented in Subsection \ref{subsec:LoRaWAN}, the maximum duty cycle and link capacity are regulated under LoRaWAN specification. Therefore, sensors in MLoRa-SS are not able to exchange information periodically as in WSN-MS. This prevents nodes from obtaining necessary information of one-hop neighbours with sufficient frequency. Consequently, when computing CA-ETX, historical variance $\sigma$ and average $\mu$ are likely to be outdated, which can significantly degrade overall routing and scheduling performance.

\end{enumerate}

Thus, a new network metric that copes with sparsity of information and node mobility is needed.

\section{Opportunistic Data Forwarding with RCA-ETX}
\label{sec:RCA-ETX}

Contact-Aware ETX (CA-ETX) has been proven effective in Wireless Sensors systems (WSN-MS), which are similar but very different as they consist of \emph{static} sensors and \textit{mobile} sinks only \cite{yang2017practical}. Although CA-ETX is not directly applicable in MLoRa-SS, we were inspired by CA-ETX and propose an approach based on a new metric called Real-time CA-ETX (RCA-ETX). This solution overcomes the uncertain nature of MLoRa-SS while inheriting the simplicity and effectiveness of CA-ETX when being used with protocols such as RPL, which base their scheduling decisions on a gradient using an objective function.

\subsection{Operation Overview of Data Forwarding with RCA-ETX}
\label{subsec:operation_overview}

LoRaWAN operates in the unlicensed spectrum and as such must adhere to the EU duty-cycling regulations. This means that devices in MLoRa-SS cannot send messages except in their allocated time slots. This makes opportunistic routing via neighbours much more challenging.
The only permitted way of finding a neighbour to route through opportunistically is via overhearing. Consequently, given a device $x$, it can only find another device $y$ when receiving (i.e. overhearing) broadcast messages from $y$.

Here we first give a high-level overview of our solution. 
Given a mobile device $x$, its opportunistic neighbour is defined as $\mathcal{D}_x(t):=\{y, \forall \textbf{C}(t)_{x,y} =1\}$.
On receiving a broadcast message from an opportunistic neighbour $y\in \mathcal{D}_x(t)$ at time $t$, $x$ computes the RCA-ETX of that link $\RCAETX_{x,y}(t)$. 
Given $\RCAETX_{x,\mathcal{S}}(t_x)$ denotes the RCA-ETX of the link in between $x$ and sinks $\mathcal{S}$, if the value of $\RCAETX_{x,\mathcal{S}}(t_x)$ is larger than $\RCAETX_{y,\mathcal{S}}(t)+\RCAETX_{x,y}(t)$, $x$ tries to send the data stored in its queue to $y$; otherwise $x$ keeps holding the data until next sending opportunity, which can be either its next sending slot or on receiving another broadcast message from another nearby device.

\begin{figure}[!ht]
\begin{center}
\includegraphics*[width=1.0\linewidth]{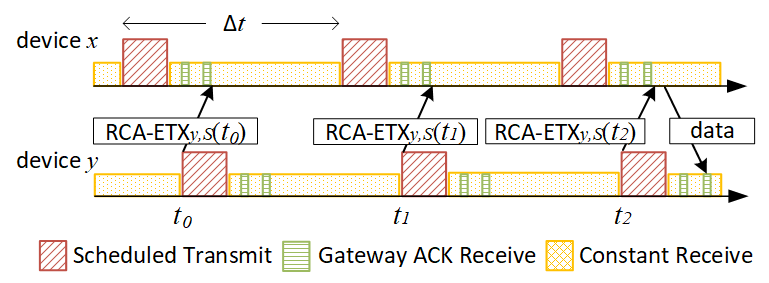}
\caption{An operational overview of our solution with LoRaWAN Class-C sensors, which are able to consistently overhear data packages from nearby neighbors). Device $x$ decides not to send its packages via $y$ when receiving its RCA-ETX sent at time $t_0$ and $t_1$, whereas $x$ sends its package to $y$ at time $t_2$.}
\label{fig:operation_verview}
\end {center}
\end{figure}

An example of such an operation is demonstrated in Figure \ref{fig:operation_verview}, where device $y$ broadcasts its $\RCAETX_{y,\mathcal{S}}(t)$ when trying to upload data to the sinks $\mathcal{S}$. 
This message can be overheard by its opportunistic neighbours $x \in \mathcal{D}_y(\cdot)$, who have no direct connection with the sinks $\mathcal{S}$ at time $t_0$, $t_1$ and $t_2$. 
On receiving $\RCAETX_{y,\mathcal{S}}(t)$ from $y$, device $x$ computes $\RCAETX_{x,y}(t)$ of link $(x,y)$ at every time $t$, and decides to handover its data to $y$ at $t_2$ after confirming that the overall RCA-ETX value of $y$ is smaller than $x$, that is:

\begin{equation}
\RCAETX_{x,\mathcal{S}}(t) > \RCAETX_{y,\mathcal{S}}(t)+\RCAETX_{x,y}(t)
\end{equation}

It is worth mentioning that, for simplicity, in the following section, \textbf{Class C} \cite{LoRaWAN} is adopted to ensure successful overhearing and device-to-device communications. Further works on reducing energy consumption will be presented in Section \ref{sec:energy_saving}.

\subsection{Calculate $\RCAETX_{x,\mathcal{S}}(t)$}
\label{subsec: node-to-sink_RPST}

Before introducing $\RCAETX_{x,\mathcal{S}}(t)$, we need to first define Packet Service Time (PST). Given a node $x$ and a set of gateways $\mathcal{S}$ in MLoRa-SS, virtual link $(x,\mathcal{S})$ can be regarded as a queue with time-varying PST $\mu_{x,\mathcal{S}}(t)\geq 0$, which is the time duration for a packet to be successfully transmitted via link $(x,\mathcal{S})$ at time $t$.
For every virtual link $(x,\mathcal{S})$ PST, $\mu_{x,S}(t)$ is computed at the beginning of every time slot reserved for its device-to-sink communication. This can be computed as
\begin{equation}
\label{eq:PST-DS}
\small
\mu_{x,\mathcal{S}}(t) = \left\{
   \begin{array}{l l}
      1/c_{x,\mathcal{S}}(t)                   &,\dot{t}_x^n \leq t \leq \ddot{t}_x^n \\
      \dot{t}_x^{n+1} - t + 1/c_{x,\mathcal{S}}(\dot{t}_x^{n+1}))              &,\text{otherwise}
   \end{array} \right.
\normalsize
\end{equation}    
where $\dot{t}_x^{n}$ and $\ddot{t}_x^{n}$ denotes the first and last time slots of $n$-th contact between $x$ and any sink $s\in \mathcal{S}$, respectively.

Here PST is regarded as the combination of transmission time $1/c_{x,\mathcal{S}}(\dot{t}_x^{n+1})$ (i.e. the PST when node $x$ in its ($n+1$)-th contact with sinks) and time delay $\dot{t}_x^{n+1} - t$ (i.e. time required before next sink contact). 
However, when computing PST, $\dot{t}_x^{n+1}$ is not available when $\ddot{t}_x^{n}<t<\dot{t}_x^{n+1}$. 
To overcome this problem, we introduce a Real-time PST (RPST) $\mu'_{x,\mathcal{S}}(t)$, which is computed as below:

\begin{equation}
\label{eq:PST-DS}
\small
\mu'_{x,\mathcal{S}}(t) = \left\{
   \begin{array}{l l}
      1/c_{x,\mathcal{S}}(t-\Delta t + t_x^\Delta) + t_x^\Delta                  &,c_{x,\mathcal{S}}(t-\Delta t + t_x^\Delta)>0 \\  
      1/c_{x,\mathcal{S}}(\ddot{t}_x^{n}) +  t - \ddot{t}_x^{n} + t_x^\Delta  &,\text{otherwise}
   \end{array} \right.
\normalsize
\end{equation}
where $\Delta t$ denotes the device-to-sink communication interval, which is a given parameter; $t_x^\Delta$ and denotes the wait time before $x$ can broadcast a message via $(x,\mathcal{S})$.

\begin{figure}[!ht]
\begin{center}
\includegraphics*[width=1.0\linewidth]{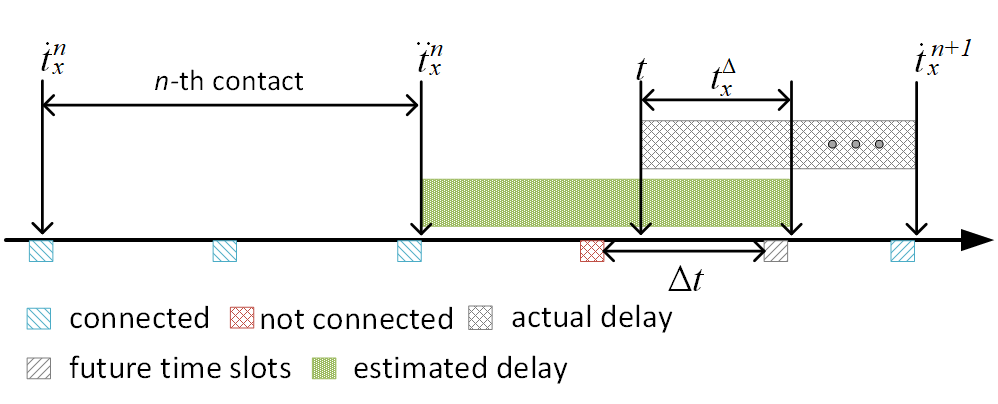}
\caption{Node-to-sink PST $\mu_{x,\mathcal{S}}(t)$ is estimated at every time $t$}
\label{fig:RCA-ETX_PST}
\end {center}
\end{figure}

Figure \ref{fig:RCA-ETX_PST} illustrates an example of computing $\mu'_{x,\mathcal{S}}(t)$.
As can be seen, given a device $x$, we use \emph{estimated delay} (i.e. $t - \ddot{t}_x^{n} + t_x^\Delta$) instead of \emph{actual delay} (i.e. $\ddot{t}_x^{n+1} - t$) that cannot be acquired. 
It is worth noting that devices use the capacity $c_{x,\mathcal{S}}(\cdot)$ acquired at the \emph{previously} successful transmission time slot $t-\Delta t + t_x^\Delta$ 
or $\ddot{t}_x^{n}$, since $c_{x,\mathcal{S}}(t)$ is also not available. 

$\RCAETX_{x,\mathcal{S}}(t)$ can now be acquired from  $\mathbb{E}[\mu'_{x,\mathcal{S}}(t)]$ at time $t$.
However, since devices are mobile and run at low-duty cycles in MLoRa-SS, the opportunity of transmission is much lower than the probability of topology changes in $\mathcal{G}$. 
Therefore, instead of the long-term average, we compute  $\mathbb{E}[\mu'_{x,\mathcal{S}}(t)]$ as exponentially weighted moving average (EWMA) as below

\begin{equation}
\label{eq:EWMA}
\small
   \mathbb{E}[\mu'_{x,\mathcal{S}}(t)] = \left\{
   \begin{array}{l l}
      (1-\alpha)\mathbb{E}[\mu'_{x,\mathcal{S}}(t-\Delta t)]+\alpha\mu'_{x,S}(t) & t>0\\
      \mu'_{x,S}(0)                                    & t=0
   \end{array} \right.
\normalsize
\end{equation}
where $\alpha$ denotes a parameter that controls weight in between current RPST $\mu'_{x,S}(t)$ and last RPST $\bar{\mu}_{x,S}(t-\Delta t)$ computed at previous time slots.
The higher the $\alpha$, the faster the RPST adapts to current RPST; however, this will reduce the overall scheduling stability. For further discussion on this parameter, we refer readers to our evaluations in Section \ref{sec:evaluation}.

\subsection{Calculate $\RCAETX_{x,y}(t)$}
\label{subsec:sensor-to-sensor_RPST}

In MLoRa-SS, each device must obey the maximum data send rate (i.e. 1\% duty cycle for most general data channels) regulated with a relatively long interval (e.g. $\Delta t=2~\text{mins}$).
As a result, given a link $(x,y)\in \mathcal{L}$, we should not use the long-term average to estimate link capacity $c_{x,y}(t)$ as in classic ETX.
Our solution here is to \emph{estimate} the link capacity as per RSSI values, which is available on receive broadcast message from other sensors. With the RSSI value $\gamma_{y,x}(t)$, $c_{x,y}(t)$ can be computed as

\begin{equation}
\label{eq:capacity}
c_{x,y}(t) = \left\{
   \begin{array}{l l l}
      c_{x,y}^\text{max}\frac{\gamma_{y,x}(t)-\gamma^\text{min}}{\gamma^\text{max}-\gamma^\text{min}}  
      & \gamma^\text{max}\geq\gamma_{y,x}(t)\geq\gamma^{\text{min}}\\
      c_{x,y}^\text{max}                                                     
      & \gamma_{y,x}(t)>\gamma^\text{max}\\
      0                                                                 
      & \gamma_{y,x}(t)<\gamma^\text{min}
   \end{array} \right.
\normalsize
\end{equation}
where $\gamma^\text{max}$ denotes the RSSI value above which $(x,y)$ should achieve its maximum capacity $c_{x,y}^\text{max}$; $\gamma^\text{min}$ RSSI value below which $(x,y)$ has minimum capacities (i.e. $0$ bits/s).

This equation is frequently used in many implementations such as the link stack in Contiki \cite{dunkels2004contiki}. 
Users may replace Equation \eqref{eq:capacity} with a hyperbolic function as per requirement; however, here we keep it simple as a proof of concept. 
Consequently, $\RCAETX_{x,y}(t)$ can be simply computed as follows:

\begin{equation}
    \RCAETX_{x,y}(t) = 1/c_{x,y}(t)
\end{equation}

\section{Real-time Opportunistic Backpressure Collection (ROBC)}
\label{sec:ROBC}

The approach proposed in Section \ref{sec:RCA-ETX} can be regarded as a greedy solution which tries to forward data via a neighbour with a shorter potential delay (i.e. smaller RCA-ETX). However, this shortest-path approach typically suffers from poor throughput performance \cite{tahir2017brpl}. Furthermore, frequent topology changes also make it less effective for finding better candidates for data handovers. 
In contrast, solutions based on queue length (e.g. BCP) have proven more effective when coping with the uncertain dynamics of mobile scenarios\cite{tahir2017brpl,yang2017practical}.

In this section, we propose another solution by integrating RCA-ETX with backpressure algorithm for MLoRa-SS.

\subsection{Queue Dynamics}
\label{subsec:queueing_dynamics}

In MLoRa-SS, we assume each device $x\in \mathcal{D}$ generates $r_x(t, \Delta t)$ amount of new data at time $t$ over the past time interval $\Delta t$. 
Each device maintains a queue (i.e. data buffer) to store data that cannot be forwarded due to no viable route to the sinks $\mathcal{S}$. 
Let $Q_x(t)\geq 0$ be the queue length of $x$ at time slot $t \geq 0$. 
Given a device $x$, from time $t-\Delta t$ to $t$, its queue length is updated as below:

\begin{equation}
\small
\label{eq:queue}
    Q_x(t)=|Q_x(t-\Delta t)-f_x^\text{out}(t,\Delta t)|_++r_x(t, \Delta t)+f_x^\text{in}(t,\Delta t)
\end{equation}
where $|\cdot|_+$ denotes an non-negative operator where given a real number $a$ (i.e. $|a|_+=a$ if $a>0$, otherwise $|a|_+=0$); $f_x^{in}(t,\Delta t)$ and $f_x^{out}(t,\Delta t)$ are the amount of total incoming and outgoing data of device $x$ in between time $t$ and $t-\Delta t$, respectively, which is defined as follows:

\begin{eqnarray}
\label{eq:in_out_data}
    f_x^\text{in}(t,\Delta t)=\sum_{y\in \mathcal{D}_x(t,\Delta t)}f_{y,x}(t,\Delta t) \\
    f_x^\text{out}(t,\Delta t)=\sum_{y\in \mathcal{D}_x(t,\Delta t)}f_{x,y}(t,\Delta t)
\end{eqnarray}
where $\mathcal{D}_x(t,\Delta t)$ denotes all $x$'s opportunistic neighbours (from which $x$ receives their RCA-ETX broadcast) over time $t-\Delta t_x$ and $t$; $f_{x,y}(t,\Delta t)$ denotes the amount of outgoing data from $x$ to $y$ over $t$ and $t-\Delta t$.

\subsection{ROCB Algorithm}
\label{subsec:ROCB_Algo}

The operation of ROBC is similar to what was proposed in Subsection \ref{subsec:operation_overview}.
Given a device $y\in \mathcal{D}$, it broadcasts its  $\RCAETX_{y,S}(t)$ and queue length $Q_y(t)$ at every $\Delta t$ interval. When a device $x$ overhears this packet, it computes its ROBC weight $\omega_{x,y}(t)$ and decides whether to pass data to $y$ or keep the data in its queue.

\subsubsection{\textbf{Weight Calculation}}
Given a device $y\in \mathcal{D}$, it broadcasts its RCA-ETX $\RCAETX_y(t)$ and queue length $Q_y(t)$ at every $\Delta t$ interval. When device $x$ overhears this packet, it computes its ROBC weight $\omega_{x,y}(t)$ as

\begin{equation}
\label{eq:ROBC_weight}
    \omega_{x,y}(t)=(Q_x(t)/\varphi_x(t)-Q_y(t)/\varphi_y(t))
\end{equation}
where
\[ \varphi_x(t)=\frac{1}{\RCAETX_{x,\mathcal{S}}(t)} \]

Here $\varphi_x(t)$ is called Real-time Gateway Quality (RGQ) of device $x$. An upper bound and a lower bound (i.e. $0<\varphi^\text{min}\leq\varphi_x\leq\varphi^\text{max}< \infty$) should be given for all sensors $x\in \mathcal{D}$ to guarantee ROBC stability \cite{yang2017practical}. 
The intuitive idea behind RGQ is similar to PST proposed in Subsection \ref{subsec: node-to-sink_RPST}. Since $\varphi_x(t)$ can be regarded as the average rate sensor a node sends its data to the sinks, it is exploited as a modifier to correct original queue lengths (i.e. how long a packet will have to wait until it is served).

\subsubsection{\textbf{Opportunistic Scheduling and Data Forwarding}}
Based on ROBC weight $\omega_{x,y}(t)$ computed in Eq. \eqref{eq:ROBC_weight}, device $x$ decides whether to handover data stored in its queue to device $y$. If $\omega_{x,y}(t)> 0$, $x$ forwards $\delta_{x,y}(t)=Q_x(t)-Q_y(t)\varphi_x/\varphi_y$ amount of data to $y$, otherwise it keeps all of its data.
This can be regarded as a scheduling problem where node $x$ decides to forward its data via an opportunistic neighbour $y$ or via itself $x$ by comparing ROBC weights $\omega_{x,y}(t)$ to $\omega_{x,x}(t)=0$.

It is worth mentioning that traditional queue-based approaches utilise full link capacity $c_{x,y}(t)$ for every transmission opportunity. However, due to the low duty cycle and the sparsity of available links in MLoRa-SS, devices only send $\delta_{x,y}(t)$ amount of data to reduce recursive loops (where data packets are sent back and forth between two devices). Therefore, device $y$ will not send data received from $x$ back even if $y$ hears from $x$ before its next forwarding opportunity to the sinks.

\section{Customised Device Classes for Practical Operations}
\label{sec:energy_saving}

\begin{figure}[!ht]
\begin{center}
\includegraphics*[width=1.0\linewidth]{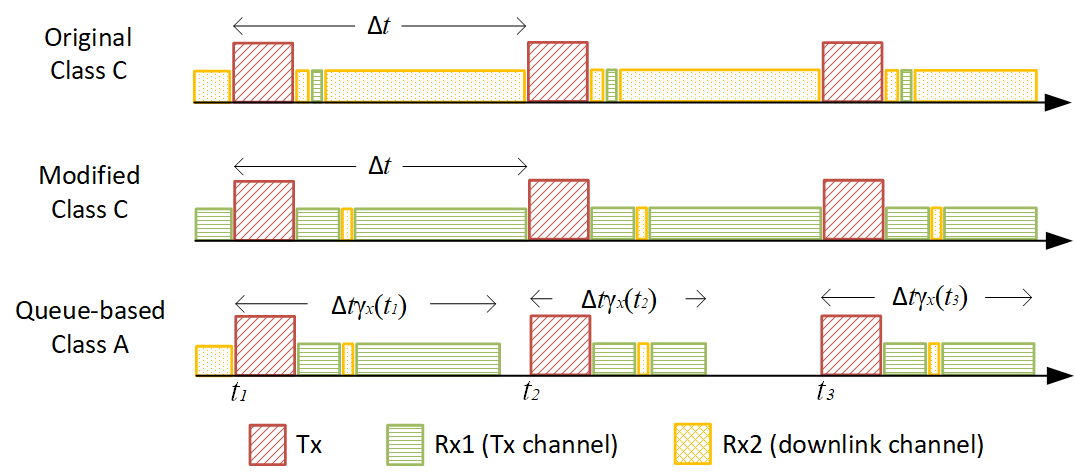}
\caption{Two new devices classes, modified Class-C and queue-based Class-A, are proposed to provide a reception feature required by device-to-device communications. \textsf{Tx} denotes the time slots reserved for device up-links. \textsf{Rx1} denotes time slots for gateway acknowledgement via the same channel as \textsf{Tx}. \textsf{Rx2} denotes the time slots for gateway down-links. }
\label{fig:Energy_Class}
\end {center}
\end{figure}

For simplicity, our solutions proposed in Section \ref{sec:RCA-ETX} and \ref{sec:ROBC} assume all devices use the \textbf{Modified Class C} extended from the specification \cite{LoRaWAN} to ensure that device-to-device communications are always available as demonstrated in Subsection \ref{subsec:LoRaWAN}. 
In addition to Modified Class C, we further enhance the practicality of our solution and propose a new queue-based Class A to accommodate adaptive duty cycling that saves energy as to its queue modified with RCA-ETX as in ROBC.
The operational overview of these two classes are demonstrated in Fig. \ref{fig:Energy_Class}.

\begin{itemize}
    \item \textbf{Modified Class C} operates in a similar manner to original Class C with one change. Instead of opening its receive slots for downlink communications from gateways (i.e. \textsf{Rx2}), it listens to its \textsf{Tx} channels (i.e. \textsf{Rx1}) to overhear the \textsf{Tx} from other nearby devices within communication range. This design is compatible with LoRaWAN Class-A, which should be supported by all LoRaWAN-enabled devices. 
    \item \textbf{Queue-based Class-A} is a dynamic approach where the length of each receive window is adapted to the size of the local data queue, which is naturally compatible with ROBC. Given a device $x$ and time $t$, the length of each receiving window is defined as $\Delta t\gamma_x(t)$, anywhere $\gamma_x(t)$ is computed from the local queue length as follows:
    \begin{equation}
    \label{eq:Gamma_Approach}
        \gamma_x(t)=\varphi_x^\text{max}Q_x(t)/\varphi_x(t)Q_x^\text{max}\leq 1
    \end{equation}
    where $Q_x^\text{max}$ is the maximum queue size of device $x$.
    The intuition behind this approach is summarised as follows:
    Since devices in MLoRa-SS are mobile and the opportunity to have opportunistic neighbours is stochastic, we argue that the longer the receiving window, the higher the probability for a sensor to receive broadcast messages from its opportunistic neighbours.
    Therefore, for the devices that have longer queues (i.e. higher $Q_x(t)$) and longer RCA-ETX (i.e. larger $\varphi_x(t)$), longer receiving slots are required to increase the probability of forwarding their data packets to the sinks. 
    Again, this class is compatible with Class-A specification of LoRaWAN.
\end{itemize}

\section{Evaluation based on London Bus Network}
\label{sec:evaluation}

In this section, we describe our experimental setup, communication network, evaluation metrics and present the results of our simulations. 

\begin{figure}[h!]
    \centering \includegraphics[width=0.48\textwidth]{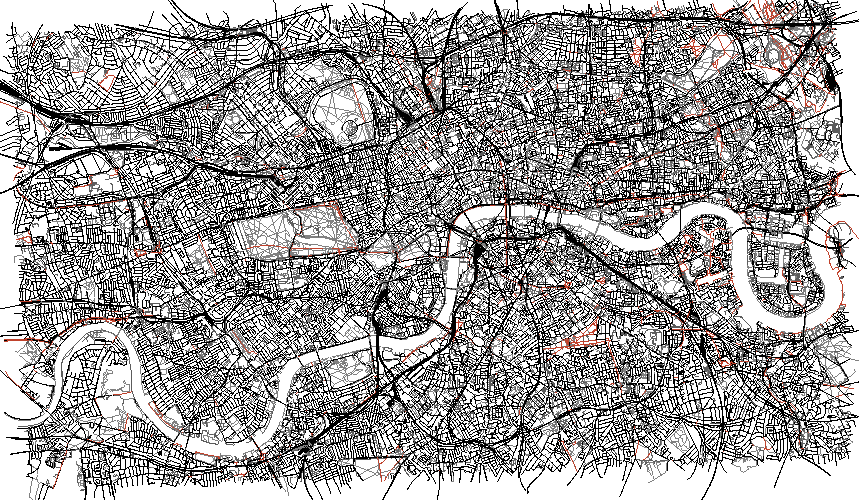}
    \caption{This map shows the simulation area where bus routes are the lines with colors.}
    \label{fig:london_bus_map}
\end{figure}

\begin{figure}[h!]
    \centering
    \begin{subfigure}[b]{0.95\linewidth}
         \includegraphics[width=\textwidth]{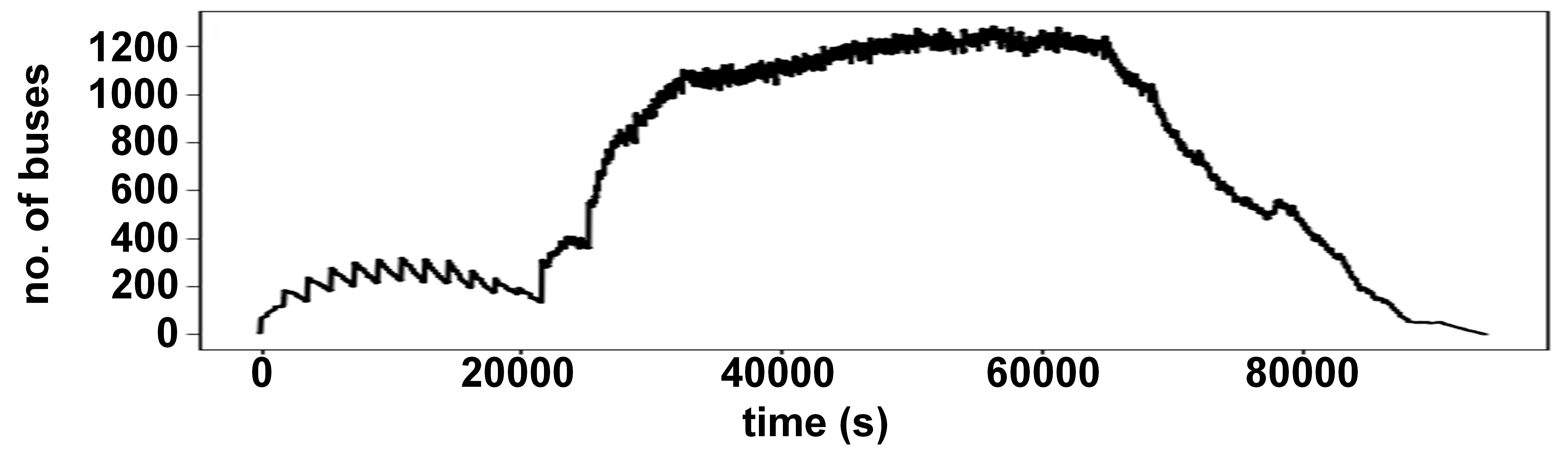}
         \caption{Number of active busses in 24 hours.}
    \end{subfigure}
    \begin{subfigure}[b]{0.95\linewidth}
         \includegraphics[width=\textwidth]{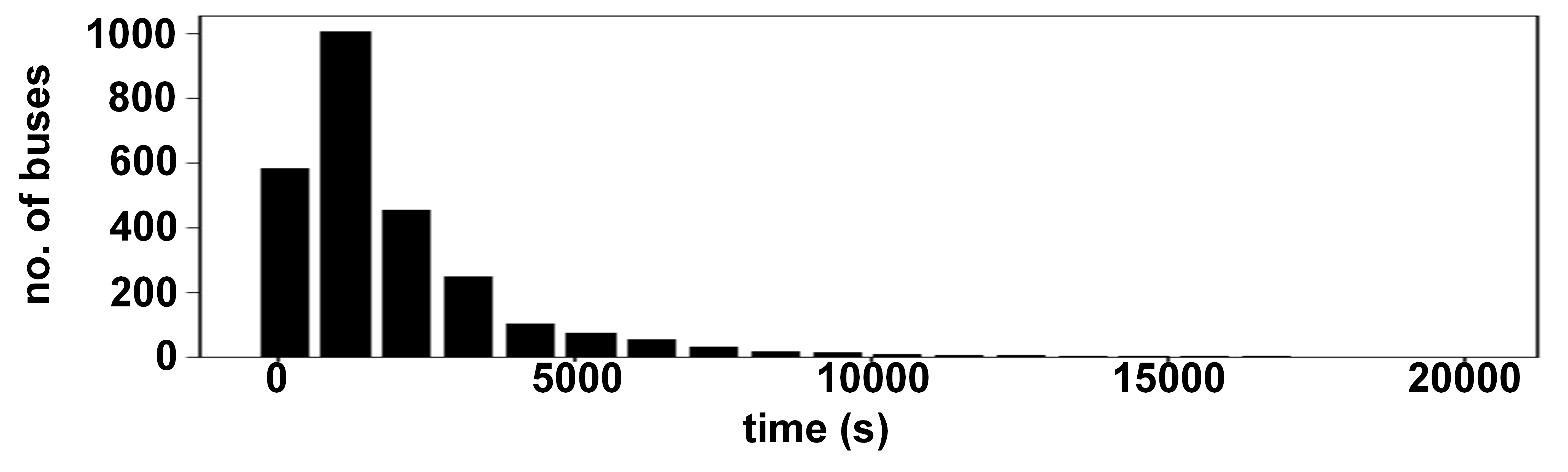}
         \caption{The distribution of bus active duration.}
    \end{subfigure}
    \caption{Here we demonstrate the property London bus networks that would potentially affect LoRaWAN communication network. This includes  (a) number of active bus and (b) number of bus active duration \cite{TFLOpen} over 24 hours.}
    \label{fig:london_bus_property}
\end{figure}

\subsection{Experimental Setup}
LoRaWAN is suitable for long-distance, low data-rate communication, and so it is particularly suitable for many urban or \emph{smart city} scenarios. The primary goal of our opportunistic data collection protocol is to collect data in areas of low coverage or over-used spectrum. One of the most suitable use-cases is logistic networks, where LoRa devices are attached to high-value parcels to track and report their conditions in real-time. 
\subsubsection{\textbf{Scenario}} To simulate logistic networks, we choose London for our evaluation as it is representative of a large-scale city. We evaluate our protocol in a 600 km$^2$ area with 24-hour simulations. 
\subsubsection{\textbf{Bus Network}} To simulate the mobility of logistic networks in such a large-scale area, we choose the trace-driven London bus network illustrated in Figure \ref{fig:london_bus_map} for our evaluation. The dynamic nature of this bus network provides an iconic example of a complex vehicle network, where the distances between buses and gateways change with the speed of each vehicle. 
For our evaluation, we collect the routes from the London bus network based on the dataset provided by Transport for London (TFL) \cite{TFLOpen}. This dataset consists of the timetable recorded from real-world for each route, based on which we can simulate the mobility of each bus.  The total number of active buses and their travel time distribution are illustrated in Figure \ref{fig:london_bus_property} (a) and (b), respectively.

\subsubsection{\textbf{Simulation Framework}} We use the SUMO simulator \cite{lopez2018microscopic}, widely used in ad-hoc traffic network research community, to create our mobility network. SUMO is a microscopic and continuous traffic simulator designed to handle large road networks. The system takes the map and bus information, like start time, stop time and frequency, to create the buses.
All the communication networks are simulated in OMNeT++ using FLORA\cite{Slabicki2018}. For every bus generated by the SUMO simulator, a corresponding node is also created in OMNeT++.

\subsubsection{\textbf{LoRaWAN Network}} 
In our simulation, LoRaWAN is a one-hop communication network where all nodes can communicate with any gateway in the network. All the gateways are connected to a single central server (network server) where all the data is collected via Ethernet.
Every bus in the simulation is equipped with a LoRaWAN device. These devices generate a 20-byte message every 3 minutes and store it in a first-in-first-out (FIFO) data queue. The devices are half-duplex, and so they are in a receive state unless they are transmitting. The nodes use modified Class-C. They are always listening on one channel which is the normal data transmission channel instead of the channel dedicated for downlink traffic. 

\subsubsection{\textbf{LoRaWAN Physical and Network Parameters}} 
In \cite{ADRMobility}, the authors prove that the benefits of Adaptive Data Rate of LoRaWAN decrease as mobility increases. For this reason, we use a single spreading factor (SF) 7 (where maximum data packet size is 255 bytes for our experiments). We also use one channel for transmission instead of multiple channels to increase the probability of overhearing a message. We use the log-distance path loss model with shadowing as the physical layer model with a path loss exponent of 2.32 as it is representative of a sub-urban environment for LoRa communications~\cite{7377400}. Upon message generation, devices select up to 12 messages from the queue and create a new data packet. They also append their RCA-ETX value and data queue size to the data packets before transmissions. 
If any gateway successfully receives the packet, it sends an acknowledgement to the devices. We assume that the acknowledgement messages are delivered instantly. If the packet fails, the nodes try to retransmit the message after the duty-cycle timer of 1\% time-on-air runs out.  Every device tries to transmit every message up-to eight times unless it generates a new packet in which case it resets the retransmission counter.
Since the nodes are all using the same SF7 and listening on the same channel, they can overhear all the messages transmitted in their neighbourhood. Upon message reception, devices can choose to forward part of their queue or ignore the reception based on the chosen data forwarding scheme.

\subsubsection{\textbf{Network Deployment}} 
The gateways are deployed in a uniform grid instead of a randomly deployed topology. Although it is highly unrealistic to have a perfect grid, it is challenging to discern the performance gain either from gateway locations or our data forwarding protocols if gateway locations are randomly chosen. 
The gateway-to-device communication range was set to 1 km for an SF7, and we use 0.5 km and 1 km to simulate device-to-device communications in an urban (where signals are likely to be blocked by buildings) and rural environments, respectively.

\subsubsection{\textbf{Schemes evaluated}} 
In this work, we evaluate three different schemes.
\begin{itemize}
    \item \textbf{NoRouting:} We use the \textbf{modified Class-C} as described in Sec.\ref{sec:energy_saving}. In addition, we introduce a queuing system at the application layer. The nodes keep all messages in queue that were not acknowledged. At every transmission, the nodes transmit a maximum of 12 messages from its queue in a data packet.
    \item \textbf{RCA-ETX:} In addition to the NoRouting scheme, the nodes also append their RCA-ETX value to the data packet. Based on the value of this metric, the nodes decide whether to forward data to another node or not.
    \item \textbf{ROBC:} In addition to the NoRouting scheme, the nodes also append their RCA-ETX value and queue lengths to the data packet. Based on the value of this metric, the nodes decide whether to forward data to another node or not. 
\end{itemize}

\begin{figure}[h!]
    \centering         
    \includegraphics[width=0.48\textwidth]{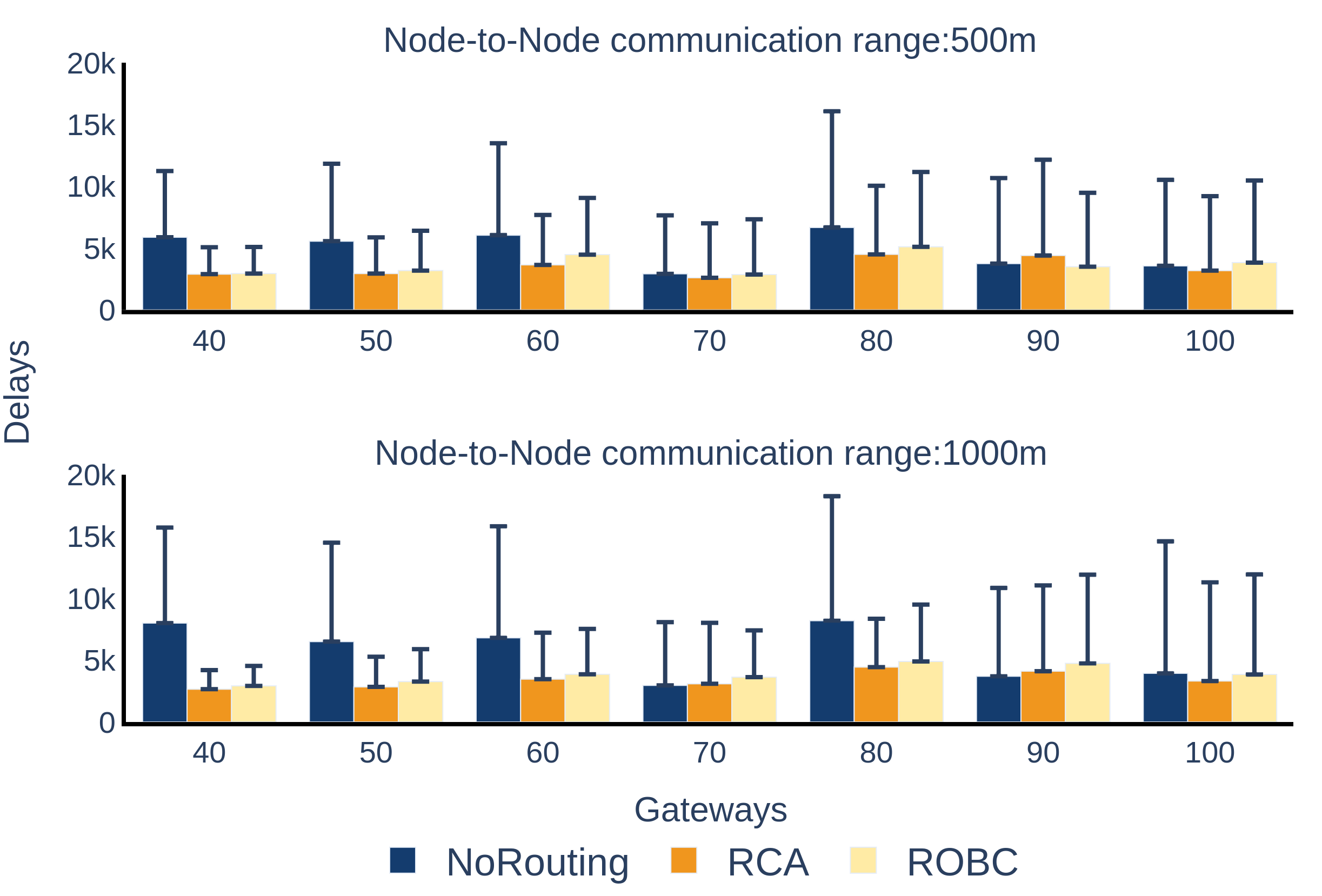}
    \caption{Average end-to-end delay with errors.}
    \label{fig:exp_delay}
    \vspace{-5mm}
\end{figure}

\begin{figure}[h!]
    \centering         
    \includegraphics[width=0.48\textwidth]{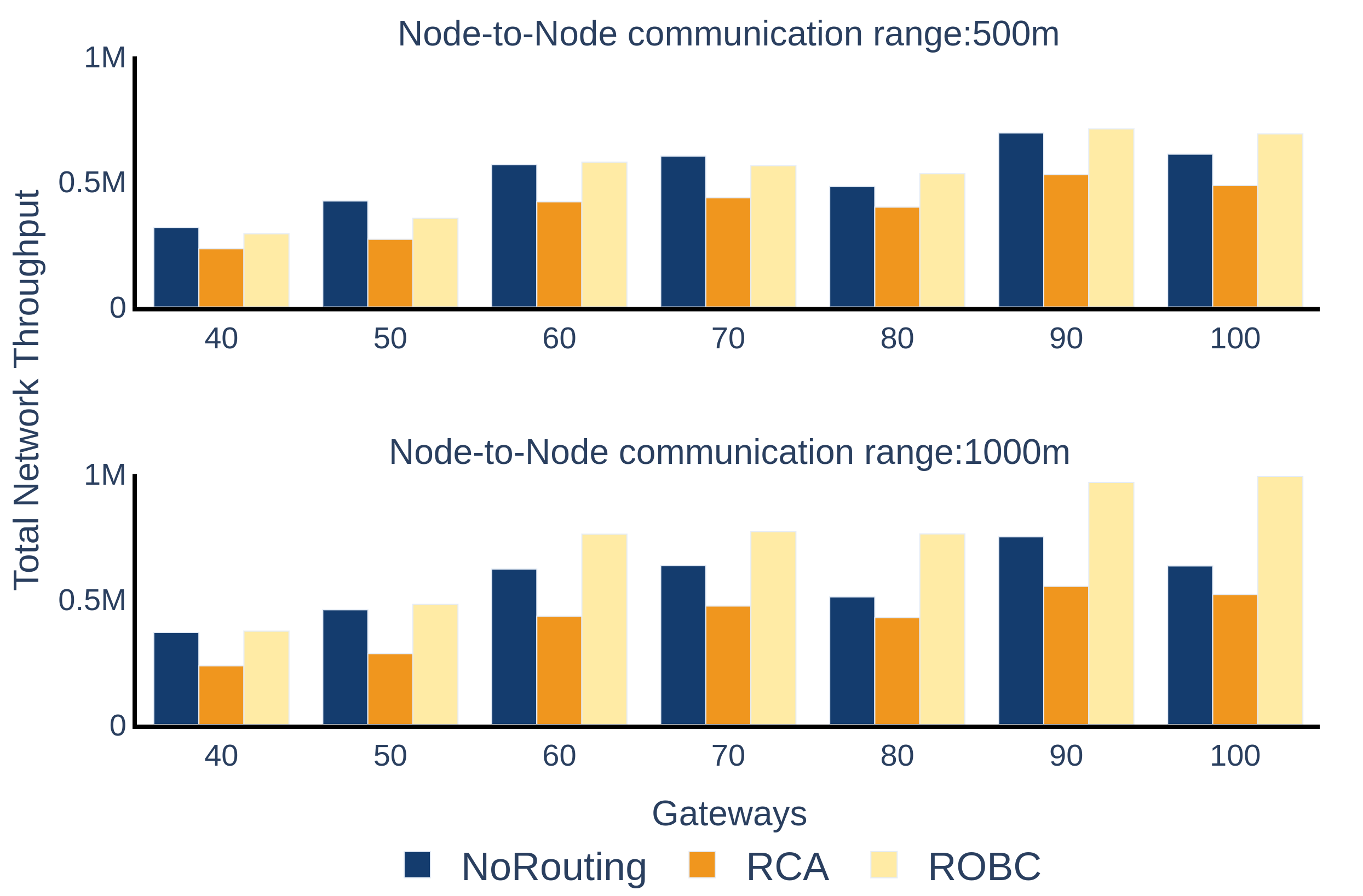}
    \caption{Total network throughput}
    \label{fig:exp_throughput}
    \vspace{-5mm}
\end{figure}

\subsection{Simulation Results}
In this subsection, we compare RCA-ETX and ROBC proposed in Section \ref{sec:RCA-ETX} and \ref{sec:ROBC}, with modified LoRaWAN without data forwarding in MLoRa-SS. The moving average parameter, that is $\alpha$ in Eq.\eqref{eq:EWMA} was set to $0.5$.
Two metrics below are used in the rest of the evaluations. 
\begin{itemize}
    \item \textbf{End-to-end delay}. One of the major contributions for our approaches is to reduce end-to-end delays by utilising nearby LoRaWAN devices for data forwarding. This end-to-end delay is measured by $\delta t(x) = t_g(x) - t_d(x)$, where $t_d(x)$  denotes the time when message $x$ is generated, and $t_g(x)$ denotes the  time when message $x$ is received at the server.
    \item \textbf{Throughput}. In contrast, throughput is a typical trade-off for the delay. This throughput is measured by the number of messages received at the server in a certain period.
\end{itemize}

\begin{figure}[ht!]
    \centering         
    \includegraphics[width=0.495\textwidth]{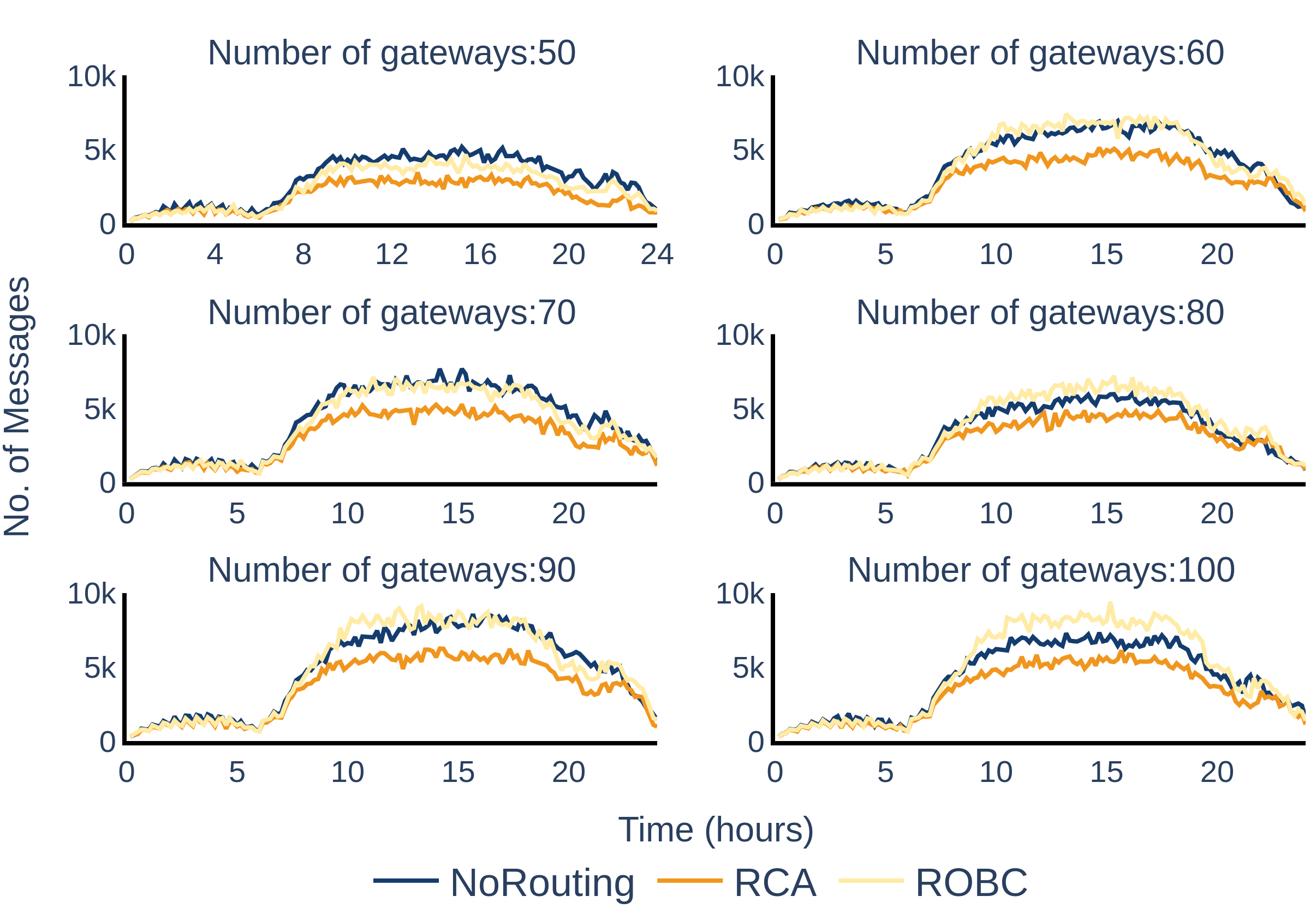}
    \caption{Throughput in urban environments where device-to-device communication range is 500m.}
    \label{fig:exp_throughput_500}
    \vspace{-5mm}
\end{figure}

\begin{figure}[ht!]
    \centering         
    \includegraphics[width=0.495\textwidth]{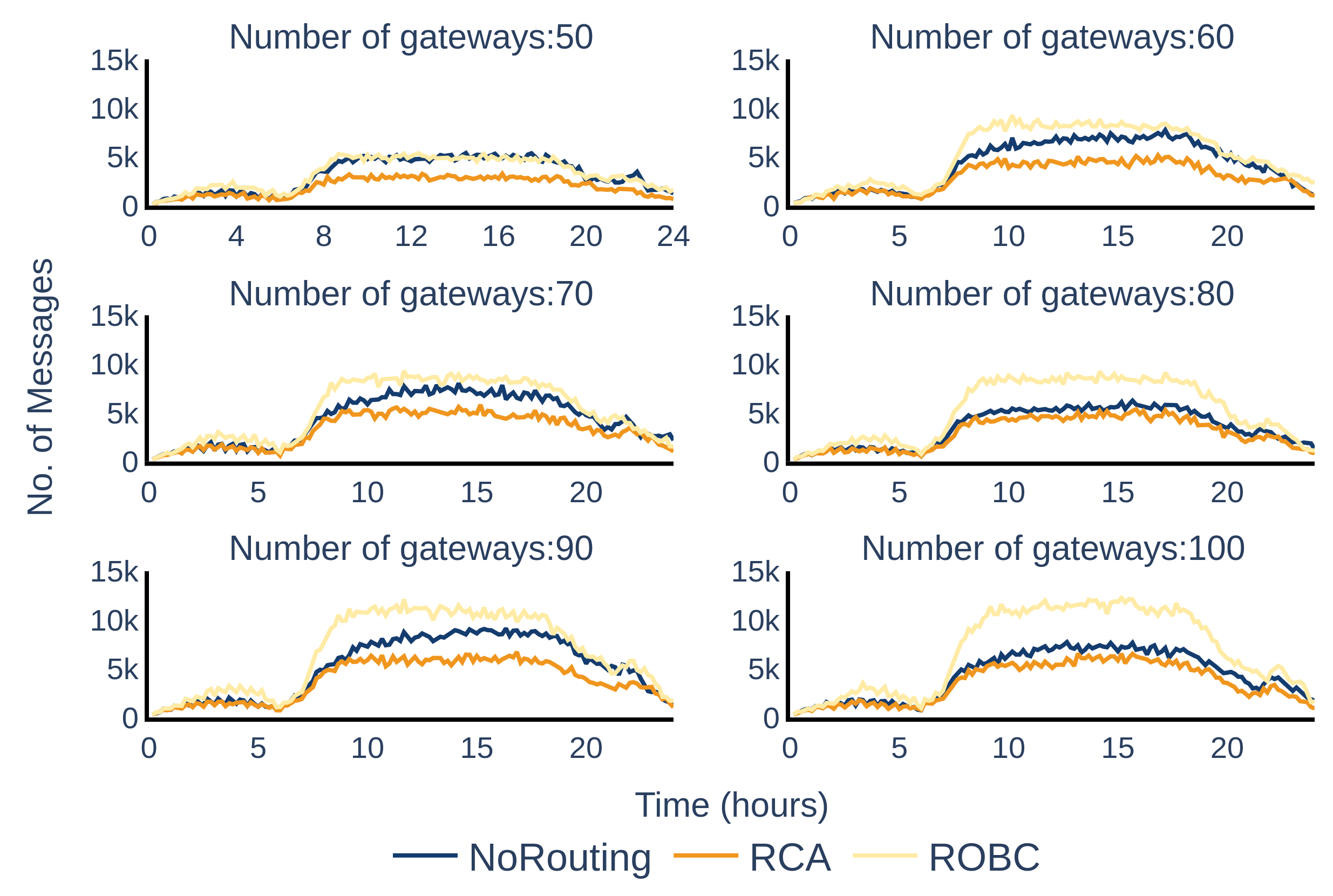}
    \caption{Throughput in rural environments where device-to-device communication range is 1000m.}
    \label{fig:exp_throughput_1000}
    \vspace{-5mm}
\end{figure}

We evaluate our approaches in networks with different gateway densities. The number of gateways deployed in the area illustrated in Figure \ref{fig:london_bus_map} is varied from 40 to 100. Also, to simulate the MLoRa-SS networks in both urban and rural environments, the device-to-device communication ranges were set to 500 m and 1000 m, respectively. Figure \ref{fig:exp_delay} depicts the average end-to-end delays. As shown, RCA-ETX and ROBC successfully reduce 10\%-25\% delays compared to original LoRaWAN without data forwarding in both urban and rural settings in lower gateway densities. However, they are less effective when gateway density increases (e.g. scenarios with 90 and 100 gateways). According to our analysis, this is because RCA-ETX is computed based on real-time PST depicted in Eq.\eqref{eq:PST-DS}. This equation exploits past information to estimate the next successful gateway contact in the future, leaving higher errors when gateways are deployed in a grid topology.
It is also worth mentioning that the performance gain in the rural environment is not as significant as expected. More extended device-to-device communication range only contributes to the scenarios where the number of gateways is less than or equal to 50.

It is worth mentioning that all three approaches yielded longer delays when the number of gateways was set to 80. Accordingly to our analysis, we realised that the density of our grid has significant impacts on gateway locations, which further impact delay and throughput. When the number of gateways equals to 80, their locations were less accessible for the buses (given their fixed route and mobility) compared to some other networks consist of fewer gateways.
However, we can still observe that RCA-ETX and ROBC still outperformed standard LoRaWAN without data forwarding.

Figure \ref{fig:exp_throughput} depicts the total throughput of the network under the same setting. As expected, RCA-ETX receives its performance gain by trading throughput. Fewer messages have arrived at the gateways. In contrast, by exploiting queue dynamics, ROBC not only reduces end-to-end delay but further improves throughput when being compared with original LoRaWAN. 
As to our observation, the improvement in throughput of ROBC is more significant when applied to rural scenarios where device-to-device communication reaches 1000~m, identical to device-to-gateway communication. The throughput increases by 38\% on average for a network with 100 gateways.

Figures \ref{fig:exp_throughput_500} and \ref{fig:exp_throughput_1000} illustrate the number of messages arriving at the gateways every 10 minutes over 24 hours. As can be seen, in urban environments, we observe RCA-ETX trades throughput for end-to-end delay performance. Meanwhile, ROBC can reach similar throughput performance as the original LoRaWAN while providing lower end-to-end delays. In rural environments, similar results are observed for RCA-ETX, while ROBC outperforms original LoRaWAN in both throughput and delay. 
Furthermore, higher bus density encourages better data forwarding between LoRaWAN devices, resulting in higher throughput when ROBC is applied. One can easily observe that higher performance gain in throughput for ROBC happens during the daytime (i.e. 20,000 - 75,000 seconds), where more buses are active as shown in Figure \ref{fig:london_bus_property} (a). The throughput achieved is up to 53\% higher compared to original LoRaWAN with 100 gateways in the rural environment. 

\begin{figure}[h!]
    \centering         
    \includegraphics[width=0.48\textwidth]{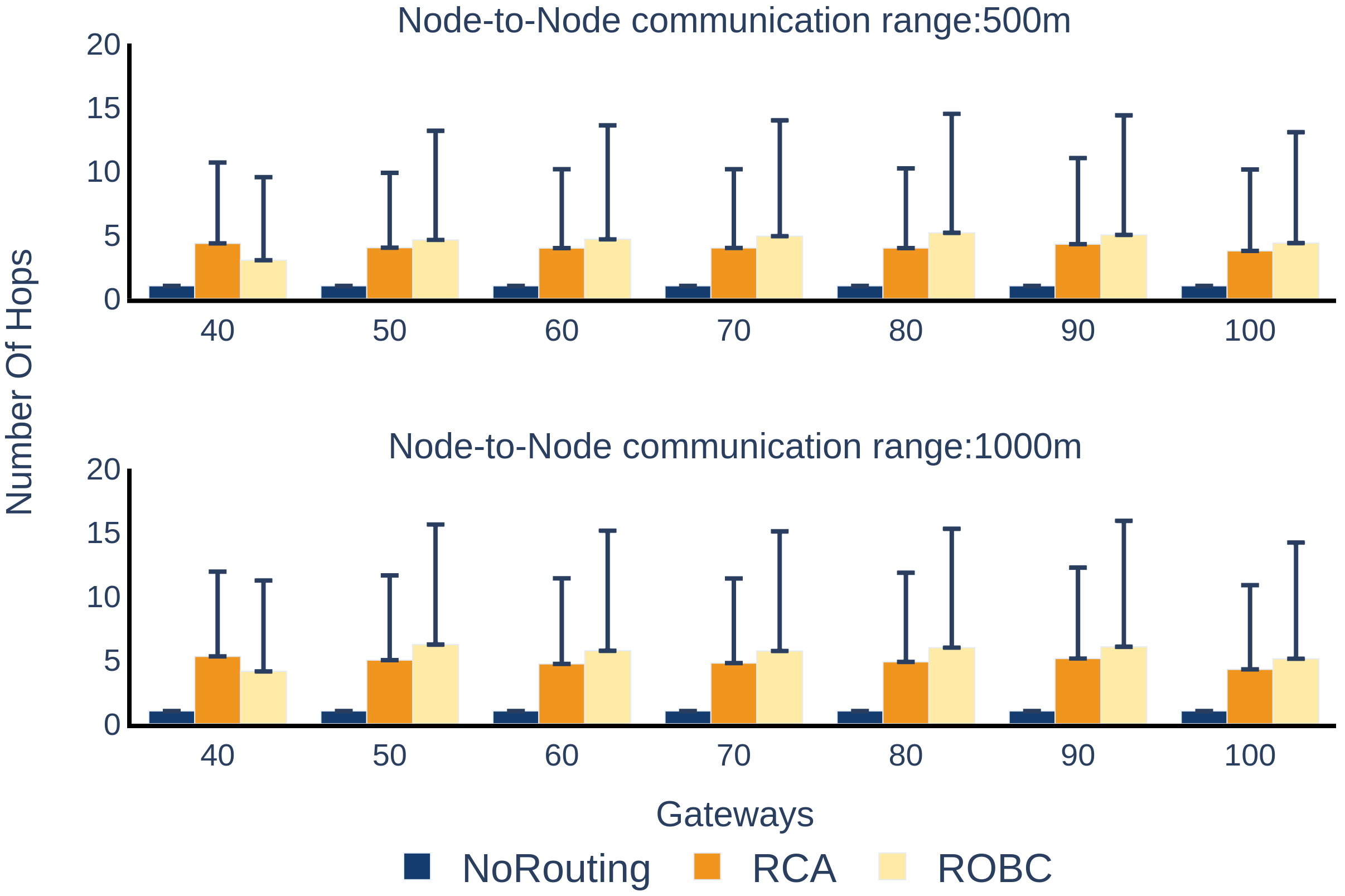}
    \caption{Average Number of hops for the entire network}
    \label{fig:numberofhops}
    \vspace{-5mm}
\end{figure}

\begin{figure}[h!]
    \centering         
    \includegraphics[width=0.48\textwidth]{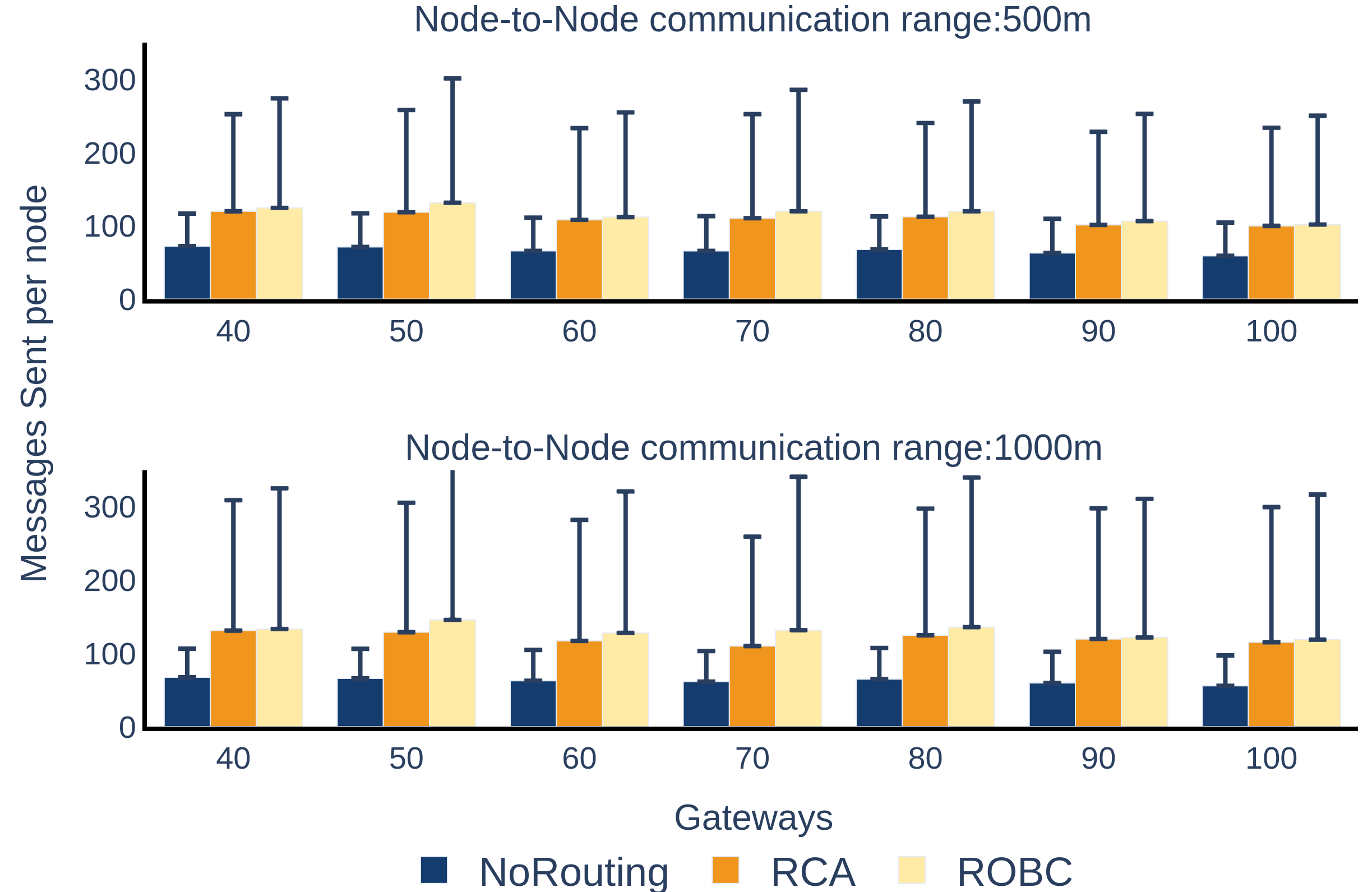}
    \caption{Average number of messages sent per node }
    \label{fig:numberofmessagessent}
    \vspace{-5mm}
\end{figure}

Figure \ref{fig:numberofhops} shows the average number of hops a message travels before reaching its destination. As can be seen in the figure, all LoRaWAN messages have a hop count of 1. The average number of hops with ROBC is higher than RCA-ETX for large number of gateways. This shows that on average a message travels 5-6 hops. The result shows that a gateway has to listen to less number of nodes at the same time as a lot of messages are bundled together before reaching the gateway.

Figure \ref{fig:numberofmessagessent} shows the average number of messages a node sends. This can approximate the energy overhead. In all our schemes, since the nodes are always listening to a particular channel to receive messages, we only consider the energy overhead as the additional number of messages sent by a node. The message overhead for RCA and ROBC is in the range of 1.6 to 2.2 times that of LoRaWAN. The network throughput can be improved by 53\% by this scheme and so the energy overhead seems reasonable as it significantly reduces the load that a gateway would need to handle.

\subsection{Further Observations and Future Work.}

Our chosen scheme uses a fixed spreading factor and one channel. As in \cite{ADRMobility}, the effectiveness of the adaptive data rate of LoRaWAN reduces as mobility increases. So, it may be better to use a single spreading factor for all buses instead of adapting it to the scenario. Our scheme is oblivious to the chosen spreading factor or channel number. As the number of spreading factors or channels used increases, the number of neighbours would reduce. We could in future optimise the system by implementing channel and SF sweeping algorithms to find neighbours using different settings.

Although we only report the results with grid-topology gateways, we have also run extensive experiments where gateways were randomly deployed. We observed that gateway locations have a significant impact on performance for these situations.
Although ROBC almost always outperforms the other two approaches, significant performance variations were also observed. Since RCA-ETX and ROBC do not have prior knowledge of the mobility and gateway locations, they both rely on \textit{estimated delay} instead of \textit{actual delay}, resulting in errors when computing RCA-ETX values. For example, comparing with a LoRaWAN device which is moving \textit{away} from gateways, a device moving \textit{towards} the gateways might have higher delays (i.e. higher RCA-ETX value) since it had less contact with gateways in the past. Consequently, selecting better gateways positioning could be another valuable research topic where we aim to find the gateway location where can better support mobility and device-to-device data forwarding in both urban and rural environments.

In this work, we also assume acknowledgements have a 100\% success ratio and gateways can acknowledge all messages. This assumption does not fit-well as gateways like end-devices have to adhere to duty-cycling. We plan to extend this work to add this duty-cyling regulation and also choosing the best gateway to acknowledge a message.

Energy is another topic worth studying in these handover systems. Our data forwarding algorithms between devices does help to reduce delay and increase throughput. However, it also requires a higher energy consumption. To this end, we implemented a Queue-based Class-A algorithm. The performance of that algorithm was on-par with the results described above, but less than 20\% energy saving was possible. So our other objective is to propose better scheduling schemes to further reduce the energy consumption.

\section{Conclusion}
\label{sec:conclusion}

We proposed a metric to provide an indicator of connectivity to support improved resilience of LoRa based mobility which we term `Real-time Contact-Aware Expected Transmission Count'. It reduces latency while improving throughput for devices in Mobile LoRaWAN Network with Static Sinks (MLoRa-SS).
We also proposed alternative routing through real-time opportunistic backpressure collection to handle the stochastic behaviours in MLoRa-SS.
We performed experiments on the London bus network to demonstrate the relevance of our work to \emph{smart city} applications. We showed that RCA-ETX could effectively reduce delays by up to 25\% by trading throughput. 
Meanwhile, ROBC can provide similar delay reductions with 15\% to maximum 53\% throughput improvement against original LoRaWAN devices with up to 2.2 times message overhead.

RCA-ETX and ROBC are already contact-aware, accounting for prior communications with base stations in decision making. For future work, we will extend this data forwarding problem to determine better communications parameters for the LoRaWAN specification (e.g. spreading factors) to improve energy usage, important for LoRaWAN devices with limited battery capacity. We found that the selection of gateway locations has a significant impact on data transfer performance, with or without data forwarding in mobile scenarios. How to find better gateway locations that optimise data transfer performance is another topic worth studying.
\balance{}
\bibliographystyle{unsrt}
\bibliography{sample.bib}

\end{document}